\documentclass[aps,prb,groupedaddress,longbibliography,preprint]{utils/revtex4-1}

\usepackage{graphicx}
\usepackage{xcolor}
\usepackage{subfiles}
\usepackage{mathbbol}
\usepackage{array}
\usepackage{tikz}
\usetikzlibrary{patterns}
\usepackage{MnSymbol}
\usepackage{pgfplots}
\pgfplotsset{compat=1.14} 
\usetikzlibrary{arrows.meta}
\usepgfplotslibrary{groupplots}

\usetikzlibrary{external}

\usepackage{mathrsfs}

\newcommand{\vect}[1]{\mathbf{#1}}

\newcommand{\fref}[1]{Figure \ref{#1}} 

\newcommand{\be}{\begin{equation}}
\newcommand{\ee}{\end{equation}}

\newcommand{\beq}{\begin{eqnarray}}
\newcommand{\eeq}{\end{eqnarray}}
\newcommand{\beml}{\begin{multline}}
\newcommand{\eeml}{\end{multline}}
\newcommand{\ba}{\begin{align}}
\newcommand{\ea}{\end{align}}

\def\btheorem{\begin{theorem}} 
\def\etheorem{\end{theorem}}
\def\blemma{\begin{lemma}}
\def\elemma{\end{lemma}}
\def\bproposition{\begin{proposition}}
\def\eproposition{\end{proposition}}
\def\bcorollary{\begin{corollary}}
\def\ecorollary{\end{corollary}}
\def\bdefinition{\begin{definition}}
\def\edefinition{\end{definition}}
\def\bexample{\begin{example}}
\def\eexample{\end{example}}
\def\bremark{\begin{remark}} 
\def\eremark{\end{remark}} 

\newlength{\boxwidth} 
\begin{document}

\title{Piezoelectric polymer generators: the large bending regime}

\author{E. Sarrey}
\affiliation{Laboratoire des Solides Irradi\'es, CEA/DRF/IRAMIS, Ecole Polytechnique, CNRS, Institut Polytechnique de Paris, F-91128, Palaiseau, France}
\author{M. Sigallon}
\affiliation{Laboratoire des Solides Irradi\'es, CEA/DRF/IRAMIS, Ecole Polytechnique, CNRS, Institut Polytechnique de Paris, F-91128, Palaiseau, France}
\author{M.-C. Clochard}
\affiliation{Laboratoire des Solides Irradi\'es, CEA/DRF/IRAMIS, Ecole Polytechnique, CNRS, Institut Polytechnique de Paris, F-91128, Palaiseau, France}
\author{A.-L. Hamon}
\affiliation{Laboratoire de M\'ecanique Paris-Saclay, Universit\'e Paris-Saclay, CentraleSup\'elec, ENS Paris-Saclay, CNRS F-91190, Gif-sur-Yvette, France}       
\author{J.-E. Wegrowe} \email{jean-eric.wegrowe@polytechnique.edu}
\affiliation{Laboratoire des Solides Irradi\'es, CEA/DRF/IRAMIS, Ecole Polytechnique, CNRS, Institut Polytechnique de Paris, F-91128, Palaiseau, France}
        
\date{\today}

\begin{abstract}
There is an important difference between piezoelectric polymer films and solid crystals for the application to piezoelectric generators. In the case of polymers, the optimal piezoelectric response imposes {\it a large bending regime}. Starting from the linear Curie's constitutive equations, we develop an analytical model under the assumption of the large bending regime resulting from bulge testing configuration. This model shows a specific non-linear piezoelectric response, that follows a power $2/3$ of the mechanical excitation. The piezoelectric voltage and the corresponding power are then studied experimentally as a function of the angular frequency $\omega$ of the mechanical excitation, the load resistance $R$, and the thickness of the film $\ell$. The experimental results carried out on piezoelectric PVDF films validate the model.
\end{abstract}

\maketitle

\section{Introduction}

The best way to extract power from a piezoelectric polymer film is to deform it. In contrast, the usual piezoelectric ceramics are not deformable at the macroscopic scale. As a result, the optimization of a polymer piezoelectric generator imposes to work in the large bending regime\cite{Timoshenko1959THEORYOP}, \cite{Here1973AcousticFA}. The specificity of such a nonlinear regime has been studied by {\it Mo et al.} in reference \cite{Mo_2014} for PVDF piezoelectric films in the context of energy harvesting.Their method (also used in subsequent works \cite{en17163886}), is based on the equilibrium states (energy minimization). The maximum energy that can be harvested has been given. However, analytical expressions for a time-dependent (or frequency dependent) mechanical excitations, and the corresponding electric power generated in a load circuit, are still to be derived.
 
The present article is a report of a systematic theoretical and experimental study about the large bending regime and the piezoelectric response to a  time-dependent excitation of the PVDF film, contacted to a load circuit. The analysis is performed on the basis of the linear constitutive laws of the Curie brothers.  

Due to the tensor nature of the equations, the number of parameters is huge : ten parameters in the case of transversely isotropic films (six for the elasticity tensor, two for the electric permittivity, and two for the piezoelectric coupling). Thus, the equations describe all possible experimental situations with various piezomaterials. In the case of piezoelectric generators, the constraints are reduced to a known external mechanical excitation and a known geometrical disposition of the two electrodes that forms the piezoelectric capacitor (so that the direction of the electrical field $\vect{E}$ is known). However, even within these restrictions, the number of possible configurations is still huge, and takes into account small to large deflection regimes.

In the context of energy harvesting at low frequency (non-resonant transducers) with polymer thin films, the main mechanical characteristics are the large transverse displacement (hundreds of micrometers), with respect to the thickness (a few tens of micrometers) for a free standing circular film fixed at the border. In order to derive operational analytical expressions, a ``large deflection'' model \cite{Dedkova2020} is used and applied to an experimental investigation performed on PVDF films. The analytical solutions to the problem are studied and compared systematically to the measured voltage and power, both for the frequency and the time domains. It is a common approach to study the output electric power with respect to the varied load resistance \cite{nano13061098}, \cite{NAVAL2023114330},\cite{WANG2023}, however, it is less common to tune it also with the frequency \cite{HASSANPOURAMIRI2023}. The experimental study is performed as a function of the frequency $\omega$ of the periodic mechanical excitation, the load resistance $R$ of the electric circuit, and the thickness of the piezoelectric film $\ell $. The main experimental observations are the following:

\begin{enumerate}
    \item For fixed material's parameters, the maximum power is reached when $R C \omega = 1$, where $R$ is the resistance of the circuit, $\omega$ the frequency, and $C$ is the effective capacitance (impedance matching condition).
    \item The ordinary differential equation of the first order in time is not linear in the mechanical excitation. Instead, the Fourier transform of the voltage $V_\omega$ is proportional to the Fourier transform of the pressure $p_\omega$ at the power of $2/3$.
    \item Scaling law: the ratio $V_\omega/(p_\omega^{2/3})$ of $V_{\omega}$ divided by $p_\omega$ to the power of $2/3$ is a simple function of the product $R \omega$ (and not a function of the two separate variables $R$ and $\omega$).
    \item The typical dependence of the voltage $V_{\omega}$ as a function of the thickness $\ell$ follows approximately the power law $V_\omega (\ell) \propto \ell^{-2/3}$.
\end{enumerate}

 While the first characteristic 1) is well-known (see however the discussion in reference Amiri {\it et al.} \cite{HASSANPOURAMIRI2023}), the properties 2), 3) and 4) are peculiar to the non-linear model presented in the present report.
Furthermore, the Figure of Merit of the system \cite{Figure_merit} is derived for the considered regime. 

The paper is composed as follows. After section II that presents briefly the set-up, section III exposes the large bending model, focusing on the four predictions exposed above. Section IV is a report of the experimental results that illustrate the main predictions of the model. The last section V presents a discussion with our conclusions. \\   

\section{Piezoelectric generator}

\begin{figure*}
    \centering
    \includegraphics{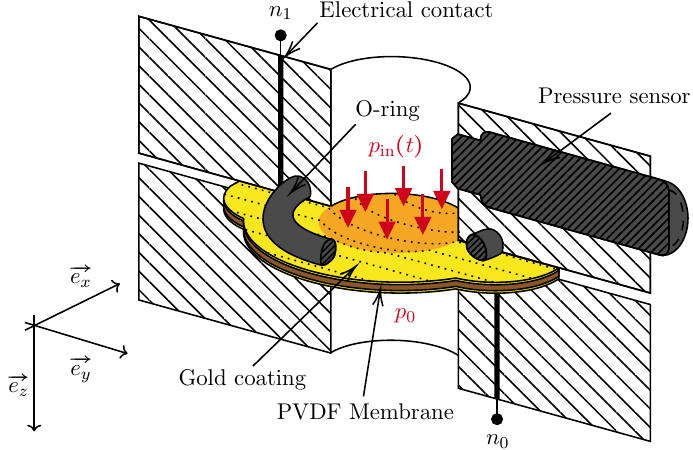}
    \caption{Illustration of the chamber with (red arrows) the applied pressure $p_\text{in}(t)$, the atmospheric pressure $p_0$, the electric connections, $n_0$ and $n_1$ and {\it in situ} pressure sensor. {\color{yellow} $\blacksquare$} the electrode area $A_e$, {\color{orange} $\blacksquare$} the pressurized area $A_p$.}
    \label{fig.setup}
\end{figure*}

The system of interest -- sketched in Fig. \ref{fig.setup} -- is composed of a piston powered by an electric motor (not depicted), with a frequency-control and connected to a pressure chamber. The pressure chamber contains the free standing PVDF film with a fixation O-ring and a pressure sensor. The film was previously covered by a sputtered gold layer (yellow surface in Fig. \ref{fig.setup}). It is contacted to the electric load circuit (top $n_1$ and bottom $n_0$ electric wires). The electric circuit is composed of a decade resistance box and a voltmeter (not shown). This configuration mimics a real autonomous piezoelectric generator, excited by mechanical vibrations at low frequency, which power is harvested with a small electronic circuit. The low frequency ($\le 32$ Hz) is chosen because it should correspond to the vibration frequencies domain of traffic-induced metallic bridge vibrations \cite{Peigney_2013}, water flows in pipelines \cite{reviewpipeline202300235}, human motion for backpack \cite{Granstrom2007} and textile sensor \cite{Nilsson2013} applications, to name but a few. \\

As justified below, the optimized system necessitates a bending of the PVDF film resulting in a bulge large with respect to the thickness (hundreds of micrometers vs. tens of micrometers), but small with respect to the lateral sizes of the piezoelectric film (centimetric). The optimized set-up corresponds to this bulge test configuration. 

In order to check the role of the large bending on the piezoelectric properties, we used three types of PVDF films with measured thicknesses $\ell$=12, $\ell$=28, and $\ell$=116$~\mu$m. The information about the PVDF films are summarized in Appendix \ref{app.spec}.

The electrodes are typically $50$~nm-thick gold layers sputtered on each side of the PVDF films. Note that the thickness of the electrode will be reconsidered in the discussion and in Appendix \ref{app.elec}. In the following, $A_e$ denotes the area of each electrode (see Fig.~\ref{fig.setup}). The PVDF film stands freely over a circular hole of $0.8$~cm in diameter, and is fixed with O-ring seals on the compression chamber (see Fig.~\ref{fig.setup}). A time-dependent pressure $p_\text{in}\left(t\right)$ of about $1$ bar in magnitude is applied with a piston. The latter motion is controlled by a motor driven by a positioning controller that imposes the time variation of the pressure as close as possible to a sinusoid. The frequency ranges from $3$ to $32$~Hz: below $3$~Hz the output piezoelectric voltage is too small, and above $32$~Hz, the signal becomes noisy due to the vibrations induced by the motor on the piston. The pressure is applied on the polymer film over a typical area $A_p$.

 The electrodes are connected to a resistance decade box of resistance $R_l$, ranging from $10^{4}$ to $10^7~\Omega$. The voltage is measured and recorded on the resistance together with the current, thanks to an analog-to-digital RedPitaya card (ADC). The effective electric circuit simply consists of the piezoelectric generator, the corresponding capacitor of capacity $C$, the load resistance $R_l$ and the $10^6 ~ \Omega$ resistance $R_{rep}$ of the ADC RedPitaya, the total resistance being $R=R_l R_{rep}/(R_l + R_{rep})$.

\section{Model}
\label{s.model}

The main objective of this section is to apply the linear Curie's constitutive laws to the geometry of our system (see Fig. 2), and the constraints imposed by our set-up.  Indeed, in the present case, the most convenient form of Curie brothers' equations is given by the {\it Stress-Charge} form \cite{IEEE_standard}
(in contrast to the more usual case based on the \emph{Strain-Charge} form):
\begin{equation}
    \begin{cases}
        \vect{T} &= \mathbb{C}_E \cdot \vect{S} - \mathbb{e}^T \cdot \vect{E} \\
        \vect{D} &= \mathbb{e} \cdot \vect{S} + \vect{\epsilon}_\text{S} \cdot \vect{E} \\
    \end{cases}
    \label{eq.S-D-tens}
\end{equation}

\begin{tabular}{lll}
    $\vect{T}$        & [N/m$^2$] & Stress            \\
    $\vect{D}$        & [C/m$^2$] & Electric charge density displacement         \\
    $\vect{S}$        & [1]       & Strain            \\
    $\vect{E}$        & [N/C]     & Electric field    \\
    $\mathbb{C}$      & [N/m$^2$] & Elasticity tensor \\
    $\mathbb{e}$      & [C/m$^2$] & Piezoelectric coupling tensor              \\
    $\vect{\epsilon}$ & [C$^2$/(N$^2$.m$^2$)] & Electric permittivity matrix \\
\end{tabular}
\\

The respective subscripts $E$ and $S$ on the elasticity $\mathbb{C}_E$ and the electric permittivity $\vect{\epsilon}_S$ mean that these quantities are respectively measured at a constant and preferably null electric field or strain.

The relation between the piezoelectric coefficient $\mathbb{e}$ and the more common Strain-Charge piezoelectric coefficient $\mathbb{d}$ is given by:
\begin{equation}
    \mathbb{e} = \mathbb{d} \cdot \mathbb{C}_E.
\label{e-d}
\end{equation}

In our specific case, taking into account the setup axisymmetric geometry and the properties of the transversely isotropic polymer film in the plane:
\begin{itemize}
    \item $\vect{T}$ and $\vect{S}$ are reduced to their radial and orthoradial components (no shear);
    \item $\vect{D}$ and $\vect{E}$ are reduced to their $z$ component,
\end{itemize}
so that the set of equations (\ref{eq.S-D-tens}) are reduced to the following set of three equations:

\begin{equation}
    \begin{aligned}
        T_{rr} &= \dfrac{Y}{1 - \nu^2} \left(S_{rr} + \nu S_{\varphi\varphi}\right) - e_{31} E_z, \\
        T_{\varphi\varphi} &= \dfrac{Y}{1 - \nu^2} \left(\nu S_{rr} + S_{\varphi\varphi}\right) - e_{31} E_z, \\
        D_z &= e_{31} \left(S_{rr} + S_{\varphi\varphi}\right) + \epsilon_{33} E_z, \\
    \end{aligned}
    \label{eqn.S-D-comp}
\end{equation}

\noindent where $Y$ and $\nu$ derive from the elasticity tensor and stand respectively for the membrane Young's modulus and Poisson's ratio, and $T$, $S$, and $D_z$ are functions of $r$, $z$, and $t$, and the electric field $E_z$ is a function of $t$ alone. From the tensor relation (\ref{e-d}) we infer the scalar relation: 
\begin{equation}
e_{31} = Y { \dfrac{d_{31} + \nu d_{33}}{\left(1 + \nu\right) \left(1 - 2 \nu\right)}}.
\label{e-d-scalar}
\end{equation}

\begin{figure*}
    \begin{center}
    \includegraphics{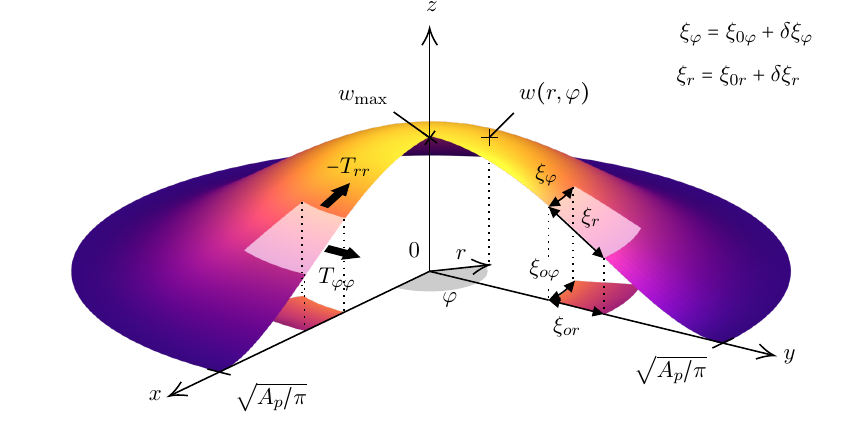}
    \end{center}
    \caption{Transformation of a circular membrane of radius $\sqrt{A_p/\pi}$ due to an overpressure $\delta p$. Cylindrical coordinates $(r,\varphi,z)$ are used. $w$ stands for the $z$-component of the displacement field for any point of polar coordinates $\left(r,\varphi\right)$, and $w_\text{max}$ its maximum value reached at the center $r = 0$. As a consequence, each radial (resp. orthoradial) segment of length $\xi_{0r}$ (resp. $\xi_{0\varphi}$) is transformed into a segment of length $\xi_r = \xi_{0r} + \delta\xi_r$ (resp. $\xi_\varphi = \xi_{0\varphi} + \delta\xi_\varphi$), so that $S_{rr} = \delta\xi_r/\xi_{0r}$ (resp. $S_{\varphi\varphi} = \delta\xi_\varphi/\xi_{0\varphi}$). The stress components $T_{rr}$ and $T_{\varphi\varphi}$ correspond to the internal force densities applied in the radial and orthoradial directions.}
    \label{fig:mecha_var}
\end{figure*}

A periodic pressure functional: $p_\text{in}\left(t\right) = p_0 + \delta p \left(t\right)$ is applied on a film of thickness $\ell$. The frequency being far smaller than the eigen-frequency of the film (a maximum of $30$~Hz versus a first resonance mode around $350$~Hz), a quasi-static approach is used for the mechanical analysis of the system.

 For a thin film such as ours, the kinematics consists in decoupling the middle plane ($z = 0$) from the thickness ($-\ell/2\le z\le\ell/2$), and each segment along the $z$-axis remains locally orthogonal to the middle plane during the transformation, so that the displacement field along the $z$-axis depends only of the plane coordinates (only $r$ in an axisymmetric problem).

In the case of small bending, the strain is uniformly null on the middle plane, so that the strain is symmetric on each side of the middle plane. From Curie's laws, the strain contribution on $D_z$ is an odd function with respect to the variable $z$, and there is no effective capacitor. Hence the need to apply an overpressure high enough for large bending $\sqrt{A_p/\pi} > w_\text{max} > \ell$ so that the strain is not a linear but an affine function of $z$. A good analytical expression of the stress in the middle plane, from which the strain can be inferred, is provided in Dedkova {\it et al.} \cite{Dedkova2020}. Adding the contribution of the $z$ variable, the strain field with respect to the position in the membrane cross-section is null for $r > \sqrt{A_p/\pi}$, and otherwise, for $r\le\sqrt{A_p/\pi}$:

\begin{equation}
    \begin{aligned}
        S_{rr}(r,z,t) &\approx S(t) \left(1 - \dfrac{1 - 3 \nu}{3 - \nu} \pi \dfrac{r^2}{A_p}\right) - z \dfrac{d^2w}{d r^2}, \\
        S_{\varphi\varphi}(r,z,t) &\approx S(t) \left(1 - \pi \dfrac{r^2}{A_p}\right) - \dfrac{z}{r} \dfrac{d w}{d r}, \\
    \end{aligned}
\end{equation}

\noindent with:
$$S\left(t\right) = \left(\dfrac{A_p}{\pi \ell^2}\right)^\frac{1}{3} \left(\dfrac{3}{8} \dfrac{1 - \nu}{7 - \nu} \dfrac{\delta p\left(t\right)}{Y}\right)^\frac{2}{3} \left(3 - \nu\right),$$

\noindent and where the $z$-component of the displacement $w$ (see Fig. 2) is given by \cite{Dedkova2020}:
$$w\left(r,t\right) = \underbrace{\left(3 \dfrac{1 - \nu}{7 - \nu} \dfrac{A_p^2}{\pi^2 \ell} \dfrac{\delta p\left(t\right)}{Y}\right)^\frac{1}{3}}_{\displaystyle w_\text{max}\left(t\right)} \left(1 - \pi \dfrac{r^2}{A_p}\right)^2.$$

The surface density of charge on the interface with each electrode ($z = \pm\ell/2$) can be thus found from eq.~(\ref{eqn.S-D-comp}), and is given in eq.~(\ref{eq.FullD}).
\begin{equation}
    \label{eq.FullD}
    D_z \left(r, \pm\dfrac{\ell}{2}, t \right) \simeq 
    \begin{cases}
        e_{31} \left[\dfrac{1 - \nu}{2} \left(\dfrac{A_p}{\pi \ell^2}\right)^\frac{1}{3} \left(3 \dfrac{1 - \nu}{7 - \nu} \dfrac{\delta p\left(t\right)}{Y}\right)^\frac{2}{3} \left(\dfrac{3 - \nu}{1 - \nu} - 2 \pi \dfrac{r^2}{A_p}\right) \mp \dfrac{\ell}{2} \left(\dfrac{d^2 w}{d r^2} + \dfrac{1}{r} \dfrac{d w}{d r}\right)\right] + \epsilon_{33} E_z \left(t\right) \\
        \text{if } r \le \sqrt{\dfrac{A_p}{\pi}} \\
    \epsilon_{33} E_z \left(t \right) \,\,\,\,\,\,\,\, \text{otherwise}.
    \end{cases}
\end{equation}

One can then take the surface average over $A_e$ (denoted $\overline{\bullet}$) and get:
\begin{equation}
\label{eq.coef}
\overline{D_z}\left(\pm\dfrac{\ell}{2},t\right) = \dfrac{A_p}{A_e} e_{31} \left(\dfrac{A_p}{\pi \ell^2}\right)^\frac{1}{3} \left(3 \dfrac{1 - \nu}{7 - \nu} \dfrac{\delta p\left(t\right)}{Y}\right)^\frac{2}{3} + \epsilon_{33} E_z\left(t\right).
\end{equation}
    
\noindent since, on average:
$$\overline{\dfrac{\partial^2 w}{\partial r^2} + \dfrac{1}{r} \dfrac{\partial w}{\partial r}} = 0.$$

Now we can use the following relations:
\be
    E_z\left(t\right) = -\dfrac{V\left(t\right)}{\ell},\qquad\text{and}\qquad\overline{\dot{D_z}} \left(t\right)= \dfrac{I\left(t\right)}{A_e},
\ee
where $V$ is the voltage, $I$ the current measured on the resistance, and $\dot{D_z} = \partial D_z/\partial t$ is the time derivative of the electric displacement. 
Eq.~(\ref{eq.coef}) can then be reduced to:
\be
   I(t) = -C \dfrac{dV}{dt} + A_p e_{31} \left(\dfrac{A_p}{\pi \ell^2}\right)^\frac{1}{3} \left(3 \dfrac{1 - \nu}{7 - \nu}\right)^\frac{2}{3} \dfrac{d}{dt} \left[\left(\dfrac{\delta p}{Y}\right)^{\frac{2}{3}}\right],
\ee
\noindent 
where we recognized the capacitance at constant or null strain:
\be\label{eq:C}
    C = \dfrac{\epsilon_{33} A_e}{\ell}.
\ee
We can then consider our full circuit where $R$ is the equivalent resistance of the load resistance coupled to measurement instrumentation. According to Ohm's law:
\be
 V = R I.
\ee
In this case, $R$ corresponds to the equivalent resistance and does not depend on the excitation frequency (assuming no other capacitance in the circuit).
Hence, we have: 
\begin{equation}
\label{ODE}
   V(t) + R C \dfrac{dV}{dt} = A_p e_{31} R \left(\dfrac{A_p}{\pi \ell^2}\right)^\frac{1}{3} \left(3 \dfrac{1 - \nu}{7 - \nu}\right)^\frac{2}{3} \dfrac{d}{dt}\left[\left(\dfrac{\delta p}{Y}\right)^\frac{2}{3}\right].
\end{equation}
Eq.~(\ref{ODE}) is an ordinary differential equation (ODE) for a driven RC circuit, for which the driven force is the time-derivative of the pressure to the power of 2/3. 
An analytical solution of eq.~(\ref{ODE}) can easily be obtained as a function of frequency after performing a Fourier transform.

\subsection{Analytical result in the frequency domain}
\label{ssec.ODE}

The solution of the ODE~(\ref{ODE}) can be more conveniently manipulated in the frequency space. Indeed, taking, the Fourier transform of eq.~(\ref{ODE}), we get:
\begin{equation}
    \hat V(\omega) \left(1 + i R C\omega \right) = \\ A_p e_{31}  R \left(\dfrac{A_p}{\pi \ell^2}\right)^\frac{1}{3} \left(3 \dfrac{1 - \nu}{7 - \nu}  \dfrac{1}{Y}\right)^\frac{2}{3} i \omega \mathscr{F}_t \left[\delta p^\frac{2}{3}\left(t\right)\right]\left(\omega\right),
\end{equation}
\noindent
with $\mathscr{F}_t$ the Fourier transform operator and $\hat V(\omega) = \mathscr{F}_t \left[V(t)\right] (\omega)$.
\noindent
To conclude, we have:
\be\label{FinalV}
    \hat V(\omega) = -\mathcal{A}(\ell) \dfrac{R \omega}{i - R C \omega} \mathscr{F}_t \left[\delta p^\frac{2}{3}\left(t\right)\right] \left(\omega\right),
\ee
\noindent
with the {\it prefactor}:
\be \label{CoefA}
    \mathcal{A}(\ell) := A_p e_{31} \left(\dfrac{A_p}{\pi \ell^2}\right)^\frac{1}{3} \left(3 \dfrac{1 - \nu}{7 - \nu} \dfrac{1}{Y}\right)^\frac{2}{3}.
\ee

The amplitude of the voltage is given by the modulus of the complex number Eq.(\ref{FinalV}):
\begin{equation}
\left\vert \hat V(\omega) \right\vert = \mathcal A(\ell) \left\vert \mathscr{F}_t \left[\delta p^\frac{2}{3}\left(t\right)\right](\omega) \right\vert \dfrac{R \omega}{\sqrt{1 + \left(R C \omega \right)^2}}.
\label{FinalVmodul}
\end{equation}
As can be seen, Eq.~(\ref{FinalVmodul}) is a scaling law showing that the voltage is a function of the product $R \omega$.

\subsection{Analytical result for the electric power delivered by the generator}
\label{ssec.power}

The apparent electric power $P_e$ delivered by the piezoelectric generator is by definition $\left\vert V\right\vert^2/(2 R)$, or
\be\label{FinalP}
    P_e\left(R,\omega\right) = \dfrac{1}2 \mathcal{A}^2(\ell) \left\vert\mathscr{F}_t \left[\delta p^\frac{2}{3}\left(t\right)\right] \left(\omega\right) \right\vert^2 \dfrac{R \omega^2 }{1 + \left(R C \omega \right)^2}
\ee

\subsection{Predictions of the model}
\label{ssec.pred}

The expression Eq.(\ref{FinalP}) is the product of three independent terms. The first is defined by the mechanical intrinsic parameters $\nu$, and $Y$, the piezoelectric parameter $e_{31}$, and by the geometrical parameters $A_p$ and $\ell$. The second is defined by the frequency dependence of the mechanical excitation only (extrinsic parameter), and the last term is a function of the load resistance $R$ (extrinsic parameter), the frequency (extrinsic parameter), and the permittivity $\epsilon$ (intrinsic parameter), through the capacitance $C$. 

The typical resonance shape of the Power, with a maximum, is due to the last term only. The maximum is reached for the condition $R C \omega = 1$. This is the maximum of the power than can be obtained for a fixed set of the internal parameters. This condition is the impedance matching between the piezoelectric generator and the electric circuit. It is an intuitive condition, since it corresponds to the situation for which the charges are injected into the electric circuit at the same rate ($\tau = RC$) as they are produced inside the piezoelectric material ($1/ \omega$). 
In other terms, the load resistance can be adapted to the typical frequency of the mechanical excitation.  

Besides, the goal is to find the best piezoelectric material for the generator. We should then look at the material parameters that maximize the prefactor $\mathcal{A}^2$ of the power Eq.(\ref{FinalP}), given in Eq.(\ref{CoefA}).  This corresponds to the maximization of the following ratio of the intrinsic parameters:
\begin{equation}
\mathcal{A}^2 \propto e_{31}^2 \, \left( 3\dfrac{1 - \nu}{7 - \nu}  \dfrac{1}{Y}\right)^\frac{4}{3},
\label{Merit}
\end{equation} 
 which defines the {\it Figure of Merit} of this generator, assuming a given mechanical input.
 
In summary, eq.~(\ref{FinalVmodul}) and eq.~(\ref{FinalP}) predict the four points listed in the introduction: 

\begin{enumerate}
    \item For fixed material's parameter, the electric power corresponds to the condition of the impedance matching: $R C \omega = 1$;
    \item The Fourier transform of the voltage is proportional to the Fourier transform of the pressure to the power of $2/3$. 
    \item  Scaling law: the ratio $V_\omega/(p_\omega^{2/3})$ is a simple function of the product $R \omega$. 
    \item Eq.~(\ref{FinalV}) and eq.~(\ref{CoefA}) show that the voltage is proportional to the thickness to the power of $-2/3$.
\end{enumerate}
The four properties above are those underlined in the introduction. The last three points are specific to the large bending regime. Note that the voltage is also proportional to the Young's modulus to the power of $-2/3$ for an input pressure modelled as a sine function (but this last feature is more difficult to evidence experimentally). 

\section{Experimental Results}

\begin{figure}[h]
    \centering
    \includegraphics{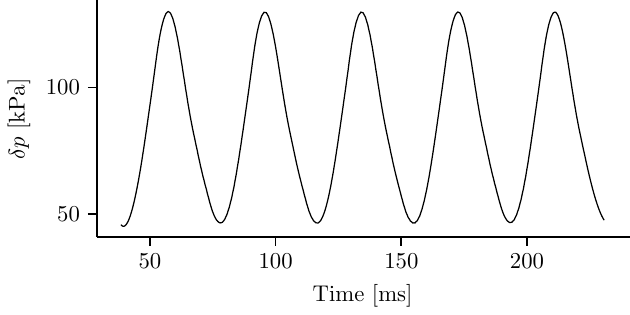}
    \includegraphics{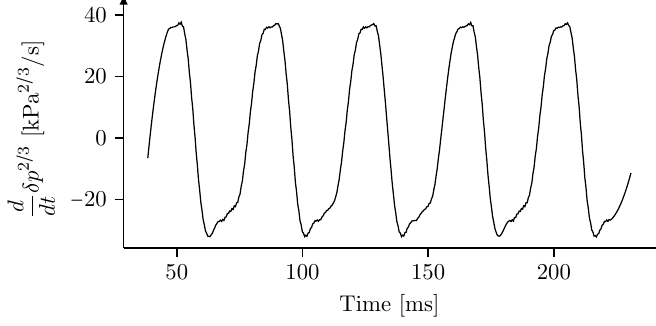}
    \caption{Top: time trace of the raw input over pressure measurement smoothed using a Savgol filter. The frequency is $26$~Hz. Bottom: derivative of the over pressure input to the power of $2/3$.}
    \label{fig:derivative}
\end{figure}

The raw data of the voltage are measured by the RedPitaya card as a function of time (time traces). The size of the trace is typically of $2^{14} = 16384$ points measured during about $1$~s (this number depends on the frequency). A typical time-trace of the pressure is shown in Fig.~\ref{fig:derivative} (top).

\begin{figure}[h]
    \centering
    \includegraphics{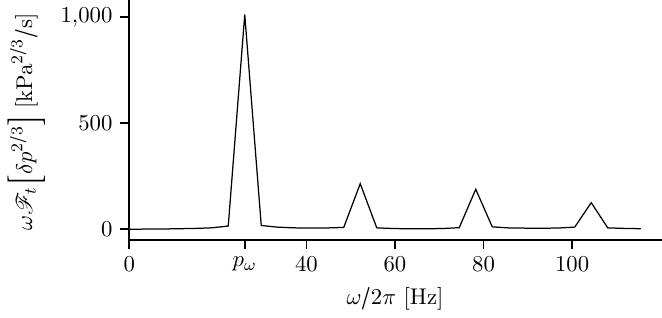}
    \caption{Numerical Fourier transform (FFT) of the time-derivative of the measured over-pressure to the power of $2/3$. The amplitude of the first peak $p_\omega$ is reached for the frequency $26$~Hz.}
    \label{fig:FTp}
\end{figure}

The voltage response trace is measured for each value of the load resistance $R_l$~-- eight values ranging from $0.01$ to $10~\text{M}\Omega$~-- and each pressure frequency~-- also eight values ranging from $8$ to $32$~Hz. An example of the raw data for the voltage is shown in Fig.~\ref{fig:trace}. The corresponding numerical time-derivative is presented in Fig.~\ref{fig:derivative} (bottom), while the numerical Fourier transform (and its time derivative) is calculated from Fig.~\ref{fig:derivative} (top) and shown in Fig.~\ref{fig:FTp}. It can be seen that the time-dependence of the pressure is nearly a sinusoidal signal  (about 93\% according to the Fourier Transform).

\subsection{Time dependence of the measured voltage}
\begin{figure}[h]
    \centering
    \includegraphics{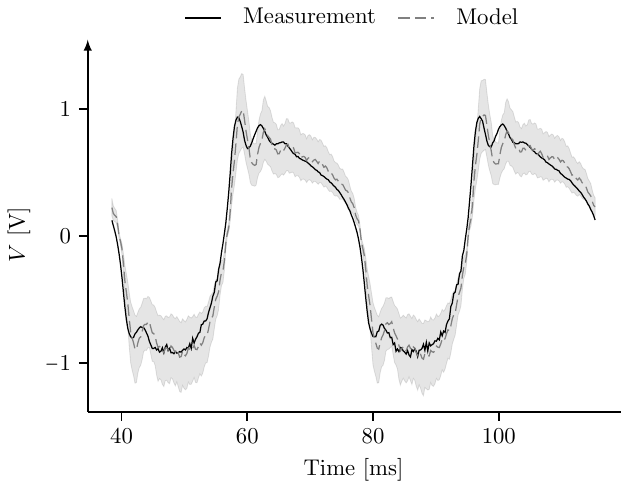}
    \caption{Time trace of the voltage over a single period. Black line: time trace of the voltage measurements. Dashed line: time trace of the model solved with Runge-Kutta with the uncertainty interval computed using the uncertainty of the material's constants given by the manufacturer. Measurement performed at 26 Hz with $909$ k$\Omega$ load on the 116$~\mu$m-thick sample.}
    \label{fig:trace}
\end{figure}

The voltage response measured on the numerical voltmeter is shown in Fig.~\ref{fig:trace} (black line). 
In order to exploit the results obtained in section \ref{s.model}, the measured pressure is derived numerically with respect to time, and used for the numerical solution (Runge-Kutta) of the differential equation (ODE) derived in \S\ref{ssec.ODE}. The corresponding function of time is plotted in Fig.~\ref{fig:trace} (dashed line) and compared to the experimental data (without fitting parameter). The grey zone accounts for the error bars deduced from the uncertainty of the material's constants given by the manufacturer.

Note that a numerical smoothing was required to prevent spikes in the derivatives which were nonphysical as they were not appearing in the voltage measurement.

As displayed in Fig.~\ref{fig:trace}, this smoothed trace can then be injected into a Runge-Kutta solver to obtain the time trace of the expected voltage for a given pressure trace.

\subsection{Frequency dependence of the measured voltage}
\begin{figure}[h]
    \centering
    \includegraphics{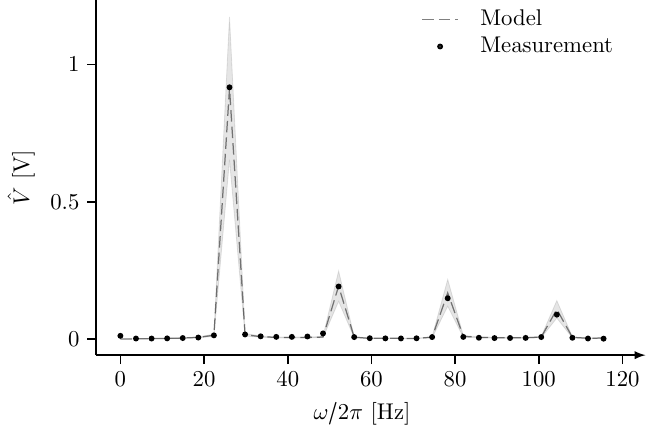}
    \caption{Comparison of the model (analytical solution) and of the voltage measurements in the frequency space. Continuous line: Fourier transform of the voltage measurements, Dashed line: analytical expression of the voltage derived in  Section III. Grey zone: error bars due to the uncertainty in the material's parameter. Measurement performed at 26 Hz with $909$ k$\Omega$ load on the 116$~\mu$m-thick sample.}
    \label{fig:fourier}
\end{figure}

The experimental frequency dependence of the voltage is studied using a numerical demodulation algorithm applied to the time-dependent measurements. \fref{fig:fourier} shows the Fourier transform of the voltage compared with the theoretical model.

\subsection{Scaling law}

\begin{figure}[h]
    \centering
    \includegraphics{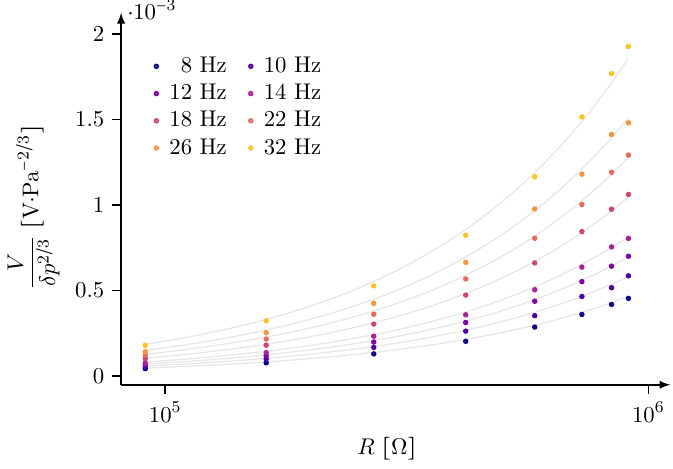}
    \caption{First peak of the Fourier transform of the voltage divided by the pressure to the power of $2/3$ as a function of the effective resistance $R$. The grey lines are fit from Eq.~(\ref{FinalVmodul}) with the fitting parameter $\mathcal A$. Measurement performed on the 116$~\mu$m-thick sample.}
    \label{fig:voltages}
\end{figure}

\begin{figure}[h]
    \centering
    \includegraphics{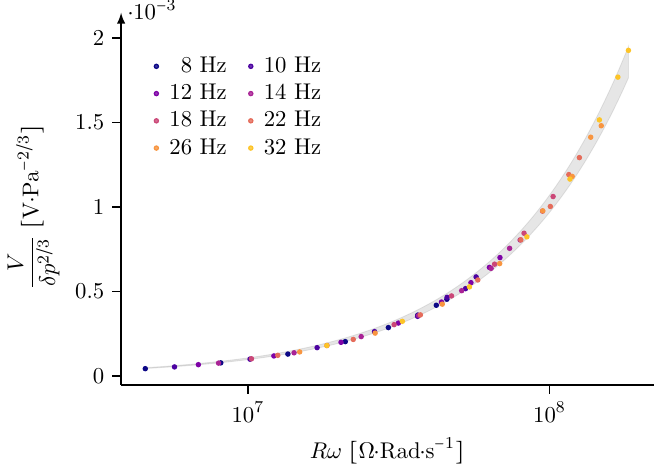}
    \caption{First peak of the Fourier transform of the voltage divided by the pressure to the power of $2/3$ as a function of the product $R \omega$. The scaling law is observed. The grey line is fit from eq.~(\ref{FinalVmodul}) with the fitting parameter $\mathcal A$. Measurement performed on the 116$~\mu$m-thick sample.}
    \label{fig:scal_law}
\end{figure}

The protocol described above is performed at multiple frequencies and multiple load resistances $R_l$. The measured the voltage divided by the Fourier transform $p_{\omega}$ of the overpressure to the power of $2/3$, plotted as a function of $R$, is shown in \fref{fig:voltages} for different excitation frequencies: changing the frequency leads to different profiles. The same values plotted as a function of the product $R \omega$ is shown in \fref{fig:scal_law}. The expected scaling law Eq.~(\ref{FinalVmodul}) is observed: all points measured at various frequencies collapse on the same curve.

This scaling law allows us to extract the value of the prefactor $\mathcal{A}$ of Eq.~(\ref{FinalVmodul}) with a single one-parameter fit for the whole set of measurements when varying both the load resistances and frequencies.

\subsection{Power}
\begin{figure}[h]
    \centering
    \includegraphics{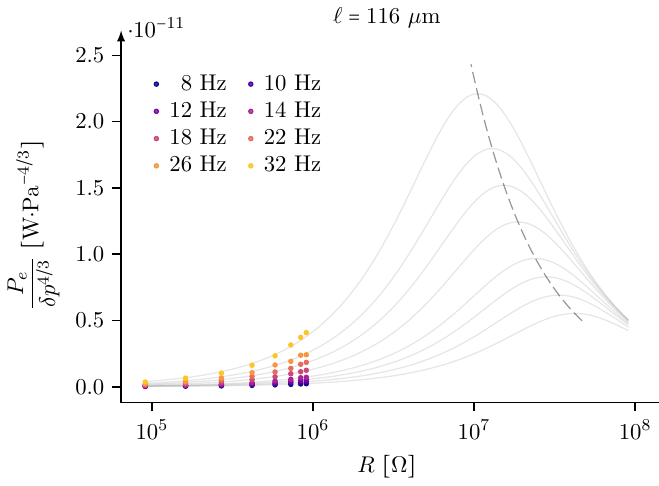}
    \vspace{1mm}
    \includegraphics{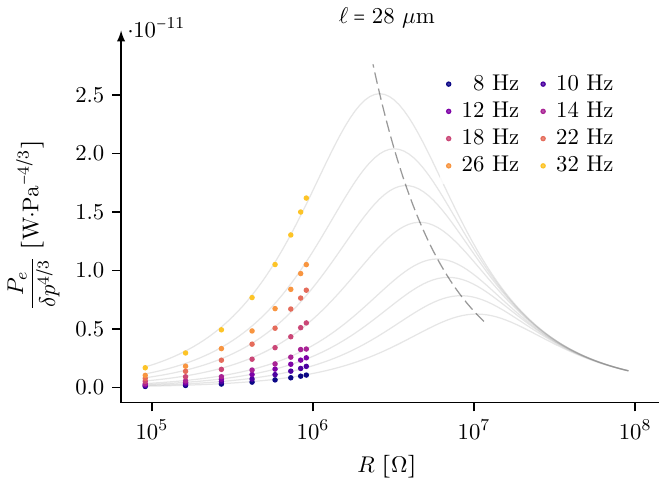}
    \caption{Measured power as a function of the effective resistance $R$ for different values of the load resistance $R_l$ and excitation frequency $\omega$. Grey dashed line: maxima at $\omega R C = 1$. Top: $\ell = 116$ $\mu$m. Bottom: $\ell = 28$ $\mu$m}
    \label{fig:Power}
\end{figure}

The power generated by the piezoelectric membrane is given by the product of the voltage by the current measured on the load resistance. The result is compared to Eq.~(\ref{FinalP}) derived in \S\ref{ssec.pred}. As for the previous subsections, the data and the model are compared as a function of the resistance, and plotted in Fig.~\ref{fig:Power} for two different thicknesses of the PVDF film ($\ell = 116$ and $\ell = 12$ $\mu$m). Such a profile can be seen on various polymer systems in the literature \cite{Jeong}. The maxima (not reached by the experimental points due to the internal resistance of the ADC card) correspond to the impedance matching $R C \omega = 1$, as predicted in \S\ref{ssec.pred}. The load resistances range within $R_{l} \le 10^7~\Omega$, but the measured points are confined to the region $R < 1$ M$\Omega$ due to the limitation of the internal resistance of our voltmeter ($1$~M$\Omega$).

\subsection{Thickness dependence}

\begin{figure}[h]
    \centering
    \includegraphics{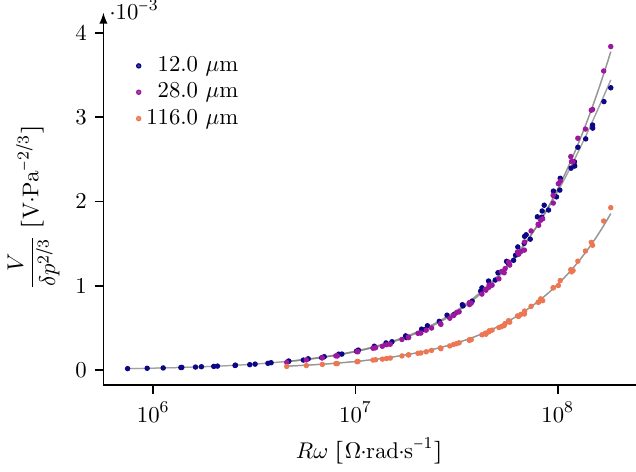}
    \caption{Scaling law for the various membrane thicknesses.}
    \label{fig:multiple_laws}
\end{figure}

Last but not least, the model can be challenged with the measurements performed on different membranes of various thicknesses, and plotted in a set of scaling laws (see \fref{fig:multiple_laws}).

\begin{figure}[h]
    \centering
    \includegraphics{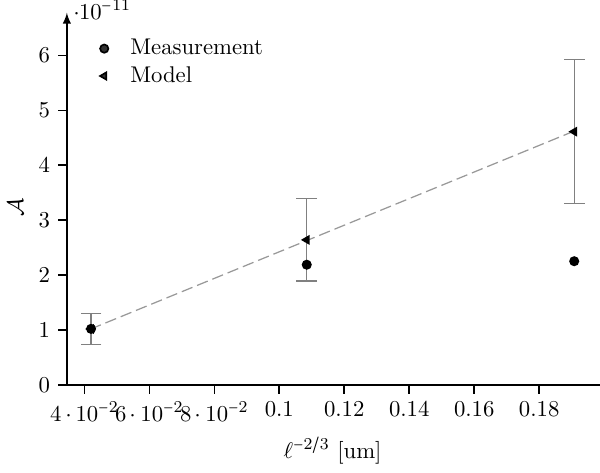}
    \caption{Prefactor $\mathcal A$ of the voltage $V$ as a function of the membrane thickness $\ell$. The deviation between the model and the data for the point corresponding to the smaller thickness $\ell= 12$  is due to the Au electrode (see text).}
    \label{fig:thickness}
\end{figure}

 For the three thicknesses $\ell = 116$, $\ell = 28$ and $\ell = 12$ $\mu$m, the scaling law is fit  as a function of the variable $R \omega$, with the parameter $\mathcal A$ as a fitting parameter. The profile of $\mathcal A$ as a function of the expected power law $\ell^{-2/3}$ (as shown in Eq.~(\ref{FinalVmodul}) is plotted in Fig.~\ref{fig:thickness}. The last value for $\ell = 12$ $\mu$m is clearly beyond the error bar. This significant discrepancy may first be attributed to the limitation of the linear constitutive equations (limit of validity of Eq.~(\ref{eq.S-D-tens})). However, this is probably not the case because we observed that the gold electrodes plays a significant role in the voltage, that accounts for the deviation, as discussed below (last section and Appendix D).  A  forthcoming study about this effect appears necessary for a correct optimization of the piezogenerators obtained from thin films.

\section{Discussion and Conclusion}

We have studied theoretically and experimentally the large bending regime of piezoelectric generators based on polymer materials, and connected to a load circuit. The polymer film is free standing with a radial deformation (it is fixed by a radial ring at the border as depicted in Fig.\ref{fig.setup}). The transverse displacement is large (several hundreds of microns), but nevertheless the linear constitutive equations of the Curie's brothers are assumed to be still valid (small strain). The large bending leads to a non-linear response of the voltage with respect to the over-pressure excitation $\delta p$. The measured power is of the order of $3 \mu$W  for a thin PVDF film of $\ell =30$ micron, of volume $10^{-9}$ $m^3$. The power density per unit of volume is of the order of $3$ $k$W/$m^3$, and the power density per unit of surface is of the order of $10$ mW/$m^2$.

An analytical model is developed in order to derive useful expressions of the Fourier transform of both the voltage and the power delivered in the load circuit.

Four main characteristics are predicted, that are 1) the well known maxima of the power imposed by the impedance matching condition ($R C \omega = 1$), 2) the $(\delta p^{2/3})_\omega$ dependence of the Fourier transform of the voltage with the over-pressure, 3) the power law dependence as a function of the product $R \omega$ of the ratio $V_\omega/(\delta p^{2/3})_\omega$, and 4) the dependence as a function of the thickness $\ell$ of the polymer film: $V_{\omega}(\ell) \propto \ell^{-2/3}$.
The verification of this power law, presented in Fig.~\ref{fig:thickness}, shows a good quantitative agreement, except for the last point at thickness $\ell = 12$ $\mu$m. This discrepancy is attributed to the unexpected role of the electrode, which is significant only for thin films. Indeed, a correction can be performed according to the empirical fit deduced from the data shown in Fig.~\ref{fig:gold_thickness}, after which the three points would be aligned in Fig.~\ref{fig:thickness}.

Besides, the efficiency of the generator can be calculated with various definitions \cite{Uchino}. Without going into the details, suffice to say that the maximum working condition is given by the relation $R C \omega = 1$ on the one hand, and on the other hand the maximum amplitude of the effect is obtained for material's parameters that maximize the product (${\mathcal A}^2 \propto e_{31}^2 Y^{-2/3})$ (see eq.~(\ref{CoefA}) and eq.~(\ref{FinalP})). This quantity, defined with intrinsic parameters of the material only, is of the order of the product of an effective piezoelectric coupling squared, $\bar{d}^2$, times Young's modulus $Y^{2/3}$, where $\bar{d} = \frac{d_{31} + \nu d_{33}}{\left(1 + \nu\right) \left(1 - 2 \nu\right)} \, \left(3 \frac{1 - \nu}{7 - \nu}\right)^{2/3}$. The product $\bar{d}^2 Y^{2/3}$ acts here as the Figure of Merit of our generator, assuming a given external mechanical excitation. It is worth pointing out that the electric permittivity, which only appears in the expression of $C$, does not participate in this Figure of Merit. More precisely, the permittivity has to be tuned independently in order to fit the maximum power condition $RC\omega = 1$.

\begin{acknowledgments}
This work was supported by the grant NanoVIBES (Labex NanoSaclay, reference: ANR-10-LABX-0035). The authors wish to specially thank Noelle Gogneau from C2N (UMR 9001) Palaiseau, France, for coordinating NanoVIBES project.
\end{acknowledgments}
\section{Conflict of Interest}
The authors have no conflicts to disclose
\section{DATA AVAILABILITY}
All data measured during this study are available from the corresponding author upon reasonable request. 
\appendix 
\section{Au-sputtered PVDF Membrane specifications}
\label{app.spec}
The 50 nm-thick gold electrodes were sputtered with a Turbomolecular pumped coater (Quorum Q150TS) on each side of the PVDF films. To adapt the geometry of the gold layers (yellow area in Fig.1) to the positions of the electric contacts in the measurement chamber, a specific home-made mask was designed. We use typically $200$ seconds duration of the deposition for 50 nm-thick gold layer. 

The piezoelectric PVDF films were purchased from GoodFellow manufacturer. 
Most of the intrinsic parameters were extracted from the manufacturer GoodFellow technical datasheet except the thicknesses and the dielectric permittivity that were respectively measured using FTIR spectroscopy and a Novocontrol broad band dielectric spectrometer. Their characteristics are: 
thickness $\ell = \{ 0.01, 0.028, 0.11\}$ $mm$,
orientation: uni-axially oriented , transparency: clear/transparent.

\begin{align*}
d_{33} & : 16 \cdot 10^{-12} \pm 10\% \, \text{C/N} \\
d_{3t} & : 3.5 \cdot 10^{-12} \pm 10\% \, \text{C/N} \\
\ell & : 12, 28 \text{ and } 116~\mu\text{m} \\
\epsilon_{33} & : 1.15 \cdot 10^{-10} \pm 10\% \text{ at 12 } \mu\text{m} \\
\epsilon_{33} & : 1.42 \cdot 10^{-10} \pm 10\% \text{ at 28 } \mu\text{m} \\
\epsilon_{33} & : 1.46 \cdot 10^{-10} \pm 10\% \text{ at 116 } \mu\text{m} \\
\text{Young's modulus } Y & : 3.2 \, \text{GPa} \\
\text{Poisson's ratio } \nu & : 0.34 \\
\text{Gold electrode area } A_{e} & : 3.75 \, \text{cm}^2 \\
\text{Pressurized area } A_{p} & : 0.5 \, \text{cm}^2 \\
\end{align*}

These parameters allow to check quantitatively the validity of the model described in section \ref{s.model}, in particular the validity of the small strain, and that the displacement amplitude remains within the range allowed by the model, as shown in Table \ref{tab.AN-model}.

\begin{table}[h!]
\begin{center}
\renewcommand{\arraystretch}{2}
\begin{tabular}{|l||c|c|c|}
\hline
thickness $\ell$\hfill[$\mu$m]                      & $12$   & $28$   & $116$  \\
\hline\hline
$\dfrac{w_{max}}{2} \sqrt{\dfrac{\pi}{A_p}}$ [$\%$] & $7.95$ & $6.00$ & $3.73$ \\
\hline
$\dfrac{w_{max}}{\ell}$\hfill[--]                   & $53.0$ & $17.1$ & $2.6$  \\
\hline
$S_{max}$\hfill[$\%$]                               & $1.68$ & $0.96$ & $0.37$ \\
\hline
\end{tabular}
\end{center}
\caption{For different membrane thicknesses $\ell$, comparison of the maximum transverse displacement and the membrane radius (row 2), or thickness (row 3), and maximum strain (row 4).}
\label{tab.AN-model}
\end{table}

\section{The effect of the electrodes}
\label{app.elec}
\begin{figure}[h!]
    \centering
    \includegraphics{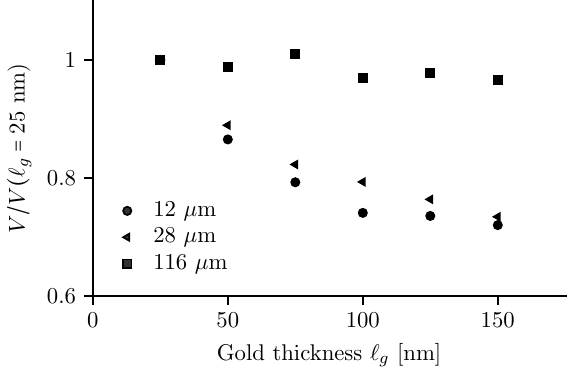}
    \caption{Effect of the gold electrodes thickness on the prefactor $\mathcal{A}$ measured by fitting the scaling law}
    \label{fig:gold_thickness}
\end{figure}

 The observation of the dependence of the piezoelectric voltage to the electrode, as shown in Fig.~\ref{fig:gold_thickness}, was a surprise. This means that the measured voltage should be corrected from the (unknown) effect of the gold electrodes, in order to take into account only the piezoelectric contribution. The correction depends on the relative thickness of the Au electrode, with respect to the thickness $\ell$ of the film. Note that the effect of the electrode thickness was already experimentally observed e.g. by Zhang et al. \cite{Zhang2013}. 

The effect of the electrode is negligible for thick films, but it becomes significant for the $12$ $\mu$m thick film. This contribution explains the quantitative deviation observed in Fig.~\ref{fig:thickness}, between the measured raw data and the model. 

\subsection{Averaged value of $\mathbf{d_{3t}}$ in the case of transverse isotropy}

As far as the specifications are concerned, the piezoelectric coupling  $\mathbb{d}$ is given by the strain-charge law:
\begin{equation}
\left\lbrace\!\!\!\begin{array}{r@{~\!\!=~\!\!}l}
\boldsymbol{S} & \mathbb{C}_E^{-1} \cdot \boldsymbol{T} + \mathbb{d}^T \cdot\boldsymbol{E} \\
\boldsymbol{D} & \mathbb{d} \cdot \boldsymbol{T} + \boldsymbol{\epsilon}_T \cdot \boldsymbol{E} \\
\end{array}\right.
\end{equation}

Nevertheless, the values provided by the manufacturer give both $d_{31}$ and $d_{32}$ components, consequence of the local texture. Yet, on our scale, the membrane is transversely isotropic, requiring to average both quantities into one single coupling factor $d_{3t}$.

For a uniaxial stress field of magnitude $T_0$ along any given direction $\mathbf{e}$ in the membrane plane $\left(\mathbf{i}_1,\mathbf{i}_2\right)$, where
$$\mathbf{e} = \cos \phi \mathbf{i}_1 + \sin \phi \mathbf{i}_2$$

\noindent the tensor expression of the stress is: 

\be
\begin{array}{r@{~= T_0~}l}
\mathbb{T} & \mathbf{e} \otimes \mathbf{e} \\
{}         & \left[\!\!\begin{array}{l}
    \cos^2 \phi \mathbf{i}_1 \otimes \mathbf{i}_1  + \sin^2 \phi \mathbf{i}_2 \otimes \mathbf{i}_2                 \\
    \qquad\qquad+ \cos \phi \sin \phi (\mathbf{i}_1 \otimes \mathbf{i}_2 + \mathbf{i}_2 \otimes \mathbf{i}_1) \\
\end{array}\!\!\right]                     \\
\end{array}
\ee

Taking into account the non null components in the case of transverse isotropy, the average effective piezoelectric coefficients are then computed from:
\be
D_z = T_0 \left(d_{31} \cos^2\phi + d_{32} \sin^2\phi\right) + \epsilon_{33} E_z
\ee

Hence, the effective piezoelectric coefficient is given by:
\begin{align}
\tilde{d}_{3t} &= \frac{1}{2\pi} \int_0^{2\pi} \left( d_{31} \cos^2 \phi + d_{32} \sin^2 \phi \right) d \phi \nonumber \\
&= \frac{d_{31} + d_{32}}{2}.
\end{align}
where $t$ refers to any in-plane direction.

\section{Direct evidence of the $\mathbf{\delta p^{2/3}}$ dependence} 
If the pressure excitation is performed with valve contacted to a pressurized tank, so that the pressure time-dependence is a step function, the maximum voltage peak follows a clear $\delta p^{3/2}$ dependence, as shown in the Fig.~\ref{fig:Fig_13_Appendix_D} below. 
The measurements were previously performed for various nanostructured piezoelectric PVDF and composite films, with e-beam irradiations of different doses with the idea to soften the materials by inducing chain scissions (as detailed in the inset for the different symbols). The study is published in the reference \cite{POTRZEBOWSKA2021102528}, however no interpretation was reported about the observed empirical power law depicting $V = f\left(\delta p\right)$. 

\begin{figure}[htp]
    \centering
    \includegraphics[width=10cm]{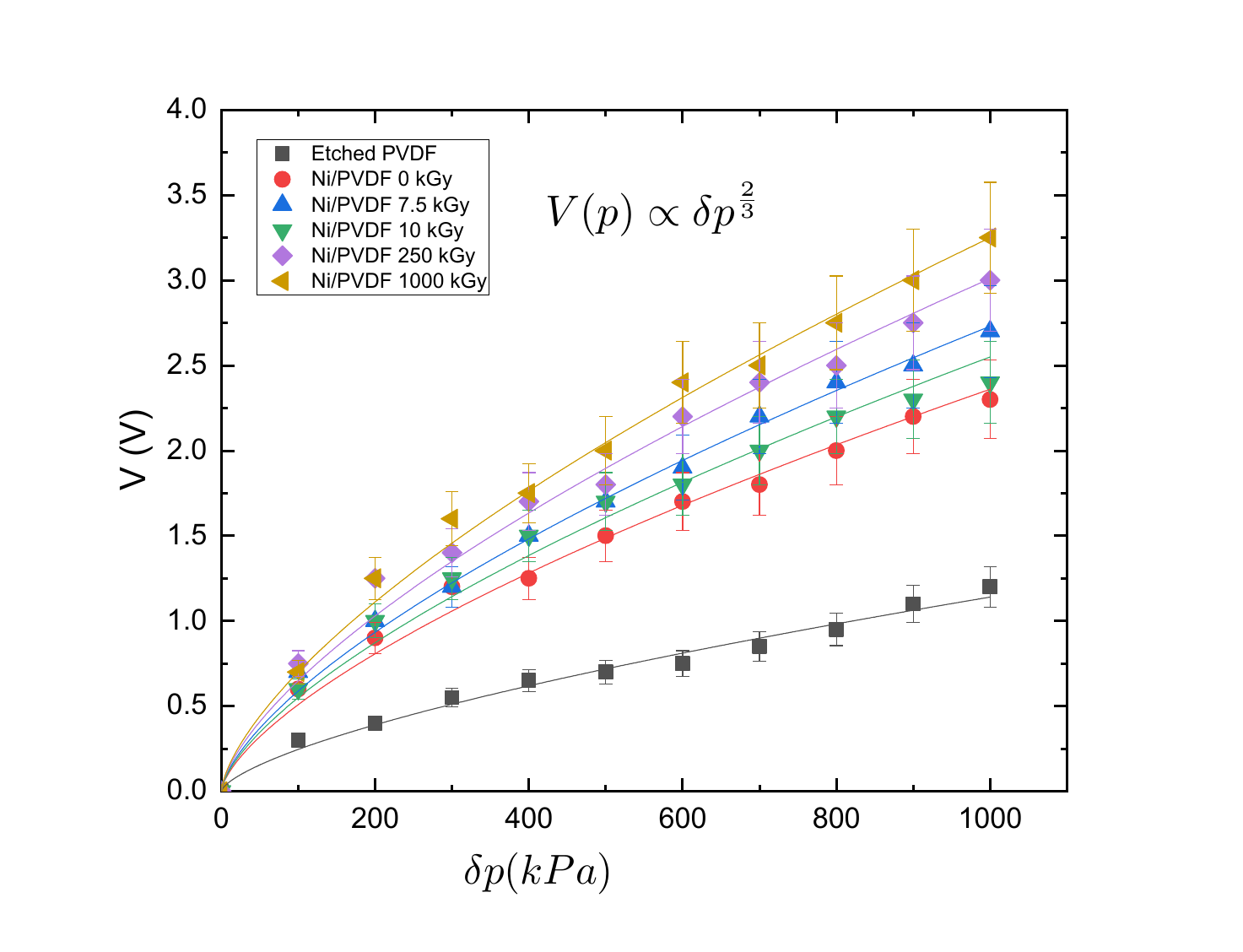}
    \caption{ Maximum piezoelectric voltage as a function of the overpressure resulting from a stepwise time-dependence excitation - raw data obtained for nanostructured piezoelectric PVDF and composite films with different irradiation treatment were extracted from reference \cite{POTRZEBOWSKA2021102528}. The lines corresponds to power laws $\delta p^{2/3}$. Measurement performed on $10$ $\mu$m thick film.\\
   }
    \label{fig:Fig_13_Appendix_D}
\end{figure}
\bibliography{bib/bibliography}
\end{document}